%% file: main.tex
\documentclass[conference,a4paper]{IEEEtran}

\usepackage{float}
\usepackage[utf8]{inputenc} 
\usepackage[T1]{fontenc}
\usepackage{url}              %
\usepackage{cite}             %

\usepackage[cmex10]{amsmath}  %
\usepackage{mleftright}       %
\mleftright                   %

\usepackage{graphicx}         %
\usepackage{booktabs}         %

\usepackage{algorithm}
\usepackage{algpseudocode} 

\usepackage[caption=false,font=footnotesize]{subfig}
\usepackage{mathtools}
\mathtoolsset{showonlyrefs=true}
\usepackage{amssymb}
\usepackage{amsmath}
\usepackage{amsthm}
\usepackage{bm}
\newtheorem{DF}{Definition}
\newtheorem{thm}{Theorem}
\newtheorem{lemma}{Lemma}
\newtheorem{ex}{Example}
\newtheorem{rem}{Remark}
\usepackage{comment}
\usepackage{multirow}

\usepackage{pgffor}
\foreach \x in {a,...,z}{%
\expandafter\xdef\csname vec\x \endcsname{\noexpand\ensuremath{\noexpand\bm{\x}}}
}
\foreach \x in {A,...,Z}{%
\expandafter\xdef\csname vec\x \endcsname{\noexpand\ensuremath{\noexpand\bm{\x}}}
}
\foreach \x in {A,...,Z}{%
\expandafter\xdef\csname c\x \endcsname{\noexpand\ensuremath{\noexpand\mathcal{\x}}}
}
\foreach \x in {A,...,Z}{%
\expandafter\xdef\csname bb\x \endcsname{\noexpand\ensuremath{\noexpand\mathbb{\x}}}
}

\newcommand{\defn}{: \, =}

\newcommand{\Lnorm}[2]{\|#1\|_{#2}}

\newcommand{\inprod}[2]{\ensuremath{\langle #1 , \, #2 \rangle}}
\newcommand{\kl}[2]{\ensuremath{D_{\mathsf{KL}}(#1\|#2)}}

\newcommand{\pr}[1]{\mathbb{P}\{ #1 \}}
\newcommand{\var}[1]{\mathsf{Var}( #1 )}

\newcommand{\Le}{\ensuremath{L(\epsilon_1,\epsilon_2,n,f)}}

\usepackage{xcolor}

\begin{document}

\title{Mean Estimation Under Heterogeneous Privacy: Some Privacy Can Be Free} 

\author{%
  \IEEEauthorblockN{Syomantak Chaudhuri and Thomas A. Courtade}
  \IEEEauthorblockA{
University of California, Berkeley
    \\\{syomantak, courtade\}@berkeley.edu}
}

\maketitle

\input{abstract}

\input{intro}
\input{results}

\input{proofs}
\input{experiments}
\input{conclusion}

\input{ack.tex}

\bibliographystyle{IEEEtran}
\bibliography{Ref}

\newpage

\appendices
\input{AppA}

\input{AppB}
\input{AppC}
\input{AppD}

\end{document}

%% file: abstract.tex
\begin{abstract}
    Differential Privacy (DP) is a well-established  framework to quantify  privacy loss incurred by any algorithm. 
    Traditional DP formulations impose a uniform privacy requirement for all  users, which is often inconsistent with real-world scenarios in which users dictate their  privacy preferences individually. 
    This work considers the problem of mean estimation under heterogeneous DP constraints, where each user can impose their own distinct privacy level.
    The algorithm we propose is shown to be minimax optimal when there are two groups of users with distinct privacy levels.
    Our results elicit an interesting saturation phenomenon that occurs as one group's privacy level is relaxed, while the other group's privacy level remains constant.  Namely, after a certain point, further relaxing the privacy requirement of the former group does not improve the performance of the minimax optimal mean estimator.  Thus,  the central server can offer a certain degree of  privacy without any sacrifice in performance.
\end{abstract}

%% file: intro.tex
\section{Introduction}

\vspace{-9pt}
Privacy-preserving techniques in data mining and statistical analysis have a long history \cite{Hoffman69,Survey00,Rao18}, and are increasingly mandated by laws such as the GDPR in Europe \cite{GDPR} and the California Consumer Privacy Act (CCPA) \cite{CCPA}.
The current de-facto standard for privacy - Differential Privacy (DP) - was proposed by \cite{DW06,Dwork06}. 
Recent extensions of DP include Renyi-DP   \cite{Mironov17}, Concentrated-DP   \cite{Dwork16}, and Zero-Concentrated-DP   \cite{Bun16}.

Statistical problems like mean estimation under privacy constraints are important in real-world applications, and there is a need to understand the trade-off between accuracy and privacy. 
Most existing  works consider a uniform privacy level for all users (see, e.g.,  \cite{Wang20}) and
do not capture the heterogeneity in privacy requirements encountered in the real-world. Such heterogeneity frequently emerges as  users balance their individual privacy options against the utility they desire from a service.
Thus, a natural question arises: how should one deal with heterogeneous privacy for statistical tasks, such as optimal mean estimation?
The effect of heterogeneity of privacy levels on accuracy is not well-understood; here, we make an effort to further the understanding of this trade-off by focusing on the mean estimation problem as a step in this direction.

We remind the readers that in the classical estimation problem without privacy constraints, the mean squared error decays as $1/n$, where $n$ is the sample size. 
While the same decay is also observed under homogeneous DP constraints,
the cost of DP is present in the second-order term, generally of form $1/(\epsilon n)^2$, where $\epsilon$ is the privacy level \cite{Cai19}.

\subsection{Our Contribution}
We consider the problem of univariate mean estimation of bounded random variables under the Central-DP model with Heterogeneous Differential Privacy (HDP). 
While the proposed scheme for mean estimation can handle arbitrary heterogeneity in the privacy levels, we prove the minimax optimality of the algorithm for the case of two groups of users with a distinct privacy constraint for each group.
This setting is particularly relevant for social media platforms, where studies have found two broad groups of users -- one with high privacy sensitivity and another that does not care \cite[Example 2]{Kotsogiannis20}.  This two-level setting is also a good first-order approximation to scenarios where users have some minimum privacy protections (e.g., ensured by legislation), but may also  opt-in   to greater privacy protections (e.g., the `do not sell my information' option mandated by the CCPA).
The setting also includes when one group's data is public, corresponding to some already known information.
For the general case of every user having a distinct privacy level, experiments confirm the superior performance of our proposed algorithm over other methods.
We view this work as a step in understanding the trade-off in the heterogeneity of privacy and accuracy; directions for further investigation are outlined in Section~\ref{sec:Con}.

Out of a total of $n$ users, a fraction $f$ are in the first group, and the rest are in the second group.
Every user in the first group has a privacy level of $\epsilon_1$ and the second group has a privacy level of $\epsilon_2$ ($\epsilon_2 \geq \epsilon_1$).
As in homogeneous DP, one might expect better accuracy in mean estimation as $\epsilon_2$ is increased keeping $n,f,\epsilon_1$ fixed\footnote{In DP framework, higher $\epsilon$ corresponds to lower privacy.}.
However, we show that after a certain critical value, increasing $\epsilon_2$ provides no further improvement in the accuracy of our estimator.
By matching upper and lower bounds, we show that this phenomenon is fundamental to the  problem and not an artifact of our algorithm.
As a corollary of this saturation phenomenon, having a public dataset ($\epsilon_2 \to \infty$) has no particular benefit for mean estimation.
Thus, the central-server can advertise and offer extra privacy up to the critical value of $\epsilon_2$ to the second group while not sacrificing the estimation performance.

We stress that our results do not assume the fraction $f$ to be constant.
For example, for a fixed $n$, one could take $f = 0$ or $f=1$ to recover results known for the homogeneous DP setting. 
One could also consider $f$ to depend on $n$; 
e.g., consider $1-f = c/n$ and $\epsilon_2 \to \infty$ to denote a constant number $c$ of public data samples as we increase the number of private samples.
The authors are unaware of any previous result that considers this problem in such a level of generality over $n,f,\epsilon_1,\epsilon_2$.
Further, many of the techniques in the literature for DP mean estimation obtain the $1/n^2$ and the $1/n$ terms  separately in the lower bound \cite{Kamath19,Duchi14}, which  cannot give tight results in $f$ that we show.

In Section~\ref{sec:PD}, we define the problem setting, state the main theorems, the proposed algorithm, along with an interpretation of the results.
Experiments and other baseline methods are presented in Section~\ref{sec:Exp} to support the theoretical claims made in this work. 
Conclusions and possible future directions are outlined in Section~\ref{sec:Con}. 

\subsection{Related Work}
Estimation error in the homogeneous DP case has been studied in great detail in recent years (see \cite{Kamath19,Hopkins22,Kamath20P,Cai19}) under both the Central-DP model and the Local-DP (LDP) model. 
In the LDP model, users do not trust the central server and send their data through a noisy channel to the server to preserve privacy \cite{Kasiviswanathan11,Duchi16}. 
Tasks like query release, estimation, learning, and optimization have been considered in the setting of a private dataset assisted by some public data \cite{Bassily20,Bassily20-learn,Liu21,Alon19,Nandi20,Kairouz21,Amid22,Wang19}. 
Using a few public samples to estimate Gaussian distributions with unknown mean and covariance matrix is considered in \cite{Kamath22}. 
The public samples eliminate the need for prior knowledge of the range of mean, but the effect on accuracy with more public samples is not considered.
HDP for federated learning is considered in \cite{Liu21Pj}. 
They remark that naively taking a linear combination of gradients in the proportion of the privacy levels is suboptimal and propose an SVD-based projected gradient algorithm.
 A general recipe for dealing with HDP is given by \cite{Alaggan17}, but their idea of scaling the data using a shrinkage matrix induces a bias in the estimator. 
 Further, their approach can not deal with public datasets.

Personalized Differential Privacy (PDP) is another term for HDP in literature. 
Reference \cite{Li17Par} studied PDP and proposed a computationally expensive way to partition users into groups with similar privacy levels. 
For each partition, standard DP algorithms can be used with respect to the minimum $\epsilon$ in each group. 
It is not immediately clear how to use these partitions for tasks like mean estimation - if we consider taking a linear combination of the outputs of each partition, then what is the optimal linear combination?
Further, for the special case of the two groups we consider, the partitions are already clearly the two groups themselves. 
An alternate method by \cite{Jorg15} proposes a mechanism that samples users with high privacy requirements with less probability. 
While this is a general approach for dealing with heterogeneity, it is not optimal for mean estimation. 
Indeed, sub-sampling when $\epsilon_2 \to \infty$
corresponds to deleting the $\epsilon_1$-private data.
Reference \cite{Ferrando21} also consider HDP mean estimation under the assumption that the variance of the unknown distribution is known.
However, as they mention, they add more noise than necessary for privacy  
since they are essentially performing LDP instead of the more powerful Central-DP technique. 
As a result, no saturation phenomenon can be deduced in their method due to the excessive noise added.
Further, they do not provide a lower bound.
The PDP setting for finite sets is consdiered in \cite{Jorg15,Niu20} and they give algorithms inspired by the Exponential Mechanism \cite{McSherry07}. 
References \cite{Li17CF,Zhang19} consider a heterogeneous privacy problem for recommendation systems.
PDP in the LDP setting has been studied by \cite{Chen16} for learning the locations of users from a finite set of possible locations.
A recent work by \cite{Cum22Mean} considered a Bayesian setting with uniform privacy and heterogeneity in the number of samples and distribution of each user’s data.
Another line of work by \cite{Torkamani22,Chatzikokolakis13} consider a more general notion of DP which encompasses HDP. 
References \cite{Avent17,Ave20,Beimel19} consider a hybrid model where some users are satisfied with the Central-DP model while other users prefer the LDP model.

Most closely related to the present work is \cite{Asu22}, which considers the general HDP setting for mean estimation in the context of efficient auction mechanism design from a Bayesian perspective. 
While they encounter a saturation-type phenomenon in their algorithm, it cannot tightly characterize the saturation condition even in the case of two datasets with distinct privacy levels (see Section~\ref{sec:Exp}).
They also assume that all the privacy levels are less than $1$. 
This assumption is central to their upper and lower bounds; hence, one cannot draw conclusions when there is a public dataset.  
Section~\ref{sec:Exp} contains more comparisons of our proposed method with that of \cite{Asu22}.

%% file: results.tex
\section{Problem Definition} \label{sec:PD}
We begin with some notation:  non-negative real numbers will be denoted by $\bbR_{\geq 0}$. 
As we consider one-dimensional data-points in our datasets, we use boldfaces, such as $\vecx$ to denote a dataset or, equivalently, a vector. 
Capital boldfaces, such as $\vecX$, denote a random dataset, i.e., a random vector. 
Vectors with subscript $i$, e.g. $\vecx_i$, refer to the $i$-th entry of the vector, while we use the notion $\vecx_i'$ for a vector differing from $\vecx$ at the $i$-th position.

A natural definition of heterogeneous Differential Privacy can be given as follows. Similar definitions were also considered in \cite{Asu22,Alaggan17}.
\begin{DF}[Heterogeneous Differential Privacy]
A randomized algorithm $M: \cX^n \to \cY$ is said to be $\bm\epsilon$-DP for $\bm\epsilon \in \bbR_{\geq 0}^n$ if 
\begin{equation} \label{eq:DP-def}
    \pr{M(\vecx) \in S} \leq e^{\epsilon_i} \pr{M(\vecx_{i}') \in S} \ \ \  \forall i \in [n],
\end{equation}
for all measurable sets $S \subseteq \cY$, where $\vecx,\vecx_i' \in \cX^n$ are any two `neighboring' datasets that differ arbitrarily in only the $i$-th component. Note that the probability is taken over the randomized algorithm conditioned on the given datasets $\vecx,\vecx_i'$, i.e., it is a conditional probability.
\end{DF}

For concreteness, we consider the case $\cX = [-0.5,0.5]$ and let $\cP$ denote the set of all distributions with support on $\cX$. 
The extension to intervals of general length is straightforward.
Under this privacy setting, we investigate the problem of estimating the sample mean from the user's data, where each user's data is sampled I.I.D. from a distribution $P \in \cP$ over $\cX$ with mean denoted by $\mu_P \in [-0.5,0.5]$ from here on. 
Each `datapoint' corresponds to a user's data in $\cX$, i.e., user $i$ has a datapoint $\vecx_i$ and the user has a privacy requirement of $\epsilon_i$.
Each user sends their data and their privacy level to the central server and the server needs to respect the users' privacy level (Central-DP model).
Let the set of all $\bm\epsilon$-DP estimators from $\cX^n$ to $\cY = [-0.5,0.5]$ be denoted by $\cM_{\bm\epsilon}$. 
We consider the error metric as Mean-Squared Error (MSE) and are interested in characterizing the minimax estimation error. 
For an algorithm $M(\cdot) \in \cM_{\bm\epsilon}$, let $E(M)$ denote the worst-case error attained by it, 
\begin{equation}
    E(M) = \sup_{P \in \cP}\ \bbE_{\vecX\sim P^n,M(\cdot)}[(M(\vecX)-\mu_P)^2].
\end{equation}
Let $L(\bm\epsilon)$ denote the minimax estimation error given by
\begin{equation} \label{eq:minimax-def}
 L(\bm\epsilon ) \defn \inf_{M \in \cM_{\bm\epsilon}}  E(M).
\end{equation}
Henceforth, we restrict our attention to the case where, out of a population of $n$ users, a fraction $f$ has a known and equal privacy requirement of $\epsilon_1$, and the rest of the population has a known and equal privacy requirement of $\epsilon_2$ %
  ($\epsilon_1 \leq \epsilon_2$ without loss of generality). 
Thus, for the described case of two groups of users, we  write $L(\epsilon_1,\epsilon_2,n,f)$ for $L(\bm\epsilon)$ defined in \eqref{eq:minimax-def}.

The notation $\gtrsim$ or  $\lesssim$ denotes inequalities that hold up to a multiplicative universal constant (independent of $n,f,\epsilon_1,\epsilon_2$).

\subsection{Main Results}

We characterize $\Le$ by giving an upper and lower bound, tight up to constant factors. 
For convenience, we define two problem-dependent quantities,
\begin{equation}
    R \defn 1 + \frac{8}{\epsilon_1^2 nf}\ ;\ \  r \defn \frac{\epsilon_2}{\epsilon_1}.
\end{equation}
We also define the averages, $\bar{\epsilon} := f\epsilon_1 + (1-f)\epsilon_2$, and $\overline{\epsilon^2} := f\epsilon_1^2 + (1-f)\epsilon_2^2$ (note $\overline{\epsilon^2}\neq \bar{\epsilon}^2$, in general). We assume $n\geq1$ throughout this work.

\begin{thm}[Upper Bound] \label{thm:UB1} 
There exists an $\bm\epsilon$-DP algorithm $M$
which attains: \\
(A) if $1 \leq r  \leq R$: 
\begin{align} \label{eq:T1A}
    E(M) &\leq \min\left\{ \frac{\overline{\epsilon^2}}{4 n \bar{\epsilon}^2} + \frac{2}{(n\bar{\epsilon})^2}, \frac{1}{4} \right\}    \\
    &= \min\left\{ \frac{fR + (1-f)r^2}{4 n [f + (1-f)r]^2}, \frac{1}{4} \right\} \ ;
\end{align}
(B) if $R \leq r$: 
\begin{align} 
    E(M) &\leq \min\left\{\frac{R}{4n[f+(1-f)R]}, \frac{1}{4} \right\}  \\
    &= \min\left\{ \frac{nf\epsilon_1^2 + 8}{4n[nf\epsilon_1^2+8(1-f)]} , \frac{1}{4}\right\} \ . \label{eq:T1B}
\end{align}

\end{thm}

The algorithm which achieves the upper bound in Theorem~\ref{thm:UB1} is outlined in Algorithm~\ref{alg:ADPM} and we refer to this algorithm as Affine Differentially-Private Mean (ADPM) in the rest of this work.
A proof sketch of Theorem~\ref{thm:UB1} is presented in Section~\ref{sec:skt-ub}. 
The weight $\vecw$ used by ADPM for the case of two groups of users is given in Table~\ref{tab:optimality}.
ADPM is inspired by the technique used by \cite{Asu22}.
Note that while we prove the optimality of ADPM for the case of two groups of users, the algorithm also works for a general $\bm\epsilon$-DP requirement. 
Even in the general $\bm\epsilon$-DP case, ADPM empirically outperforms other existing algorithms  (see Section~\ref{sec:Exp}).

\begin{thm}[Lower Bound]
\label{thm:LB1}
The minimax estimation error defined in \eqref{eq:minimax-def} satisfies:\\
(A) if $1 \leq r  \leq R$: 
\begin{align} \label{eq:T2A}
    \Le &\gtrsim \min\left\{ \frac{\overline{\epsilon^2}}{4 n \bar{\epsilon}^2} + \frac{2}{(n\bar{\epsilon})^2}, \frac{1}{4} \right\} \ ;
\end{align}
(B) if $R \leq r$:
\begin{align} \label{eq:T2B}
    \Le \gtrsim \min\left\{ \frac{R}{4 n [f + (1-f)R]}, \frac{1}{4} \right\} \ .
\end{align}
\end{thm}

A proof sketch of Theorem~\ref{thm:LB1} is given in Section~\ref{sec:skt-lb}. 
Theorem~\ref{thm:UB1} and Theorem~\ref{thm:LB1} together characterize the minimax estimation error $\Le$ up to constant factors, demonstrating optimality of ADPM (modulo universal constant factors).

Theorems \ref{thm:UB1} and \ref{thm:LB1} together demonstrate a fundamental saturation phenomenon of practical importance that occurs when  $\epsilon_2$ is large. 
In particular, for $\epsilon_2 \geq R\epsilon_1$, the  accuracy of any optimal algorithm does not further improve (modulo constant factors)  if $\epsilon_2$ is increased.
In other words, the central server gains no improvement in the accuracy of mean estimation if the group with the lower privacy level keeps lowering their privacy level after a certain point.
Thus, the central server might as well offer a privacy level of $R\epsilon_1$ to this group of users at no cost to the server.
This starkly contrasts the homogeneous-DP case, where the central server gains accuracy as the privacy level for everyone is lowered.

\begin{algorithm}
\caption{Affine Differentially Private Mean (ADPM)}
\label{alg:ADPM}
\begin{algorithmic}[]
 \Procedure{ADPM}{$\bm\epsilon,\vecx$}
\begin{align*}
\hspace{16pt} \text{Solve:}  \hspace{5pt}  \vecw^{*} = 
 \left\{
 \begin{array}{ll}
       \text{argmin} &  \frac{\Lnorm{\vecw}{2}^2}{4} + 2 \Lnorm{\vecw/\bm\epsilon}{\infty}^2 \\
       \text{subject to:} & \vecw \succcurlyeq \bm 0, \ \ \sum_{i=1}^n w_i = 1 \\
 \end{array} 
 \right.
 \end{align*}
 \Comment{$\vecw/\bm\epsilon$ is element-wise division}
 \If{$ \frac{\Lnorm{\vecw^*}{2}^2}{4} + 2 \Lnorm{\vecw^*/\bm\epsilon}{\infty}^2 > \frac{1}{4}$}
\State \textbf{return} $0$
\Else
 \State $N \sim \text{Laplace}(\Lnorm{\vecw^{*}/\bm\epsilon}{\infty})$ 
 \State \textbf{return} $\inprod{\vecw^{*}}{\vecx} + N$
 \EndIf
\EndProcedure
\end{algorithmic}
\end{algorithm}

\begin{table}[h]
\centering
\caption{Optimal weights obtained by ADPM: $w_i$ refers to the weights assigned to users of group $i$.}
\label{tab:optimality}
\begin{tabular}{ccc}
\toprule
\textbf{Condition}  &
  \textbf{Optimal \boldmath$w_1$} &
  \textbf{Optimal \boldmath$w_2$} \\ 
  \midrule 
\textbf{$\epsilon_2 \leq R\epsilon_1$} &
  $\epsilon_1/n\bar{\epsilon}$ &
  $\epsilon_2/n\bar{\epsilon}$ \\
  \hline
\textbf{$\epsilon_2 \geq R\epsilon_1$} &
  $1/n[f+(1-f)R]$ &
  $R/n[f+(1-f)R]$ \\
  \bottomrule
\end{tabular}
\vspace{-13pt}
\end{table}

\begin{rem}
From Table~\ref{tab:optimality}, it is interesting to note that if we keep other parameters constant and increase $\epsilon_2$ from $\epsilon_1$ to $\infty$, then initially, the optimal affine estimator assigns more weight to the $\epsilon_2$-dataset. This can be intuitively understood by considering that this dataset needs less privacy so we give a higher weight to it in the estimator. 
However, arbitrarily increasing $\epsilon_2$ should not increase the weight for its corresponding dataset since this comes at the cost of higher variance due to effectively ignoring the $\epsilon_1$ dataset. 
Indeed, when $\epsilon_2$ crosses the threshold of  $R \epsilon_1$, there is no further change in the weights. 
It can be understood as `saturating' the weight after this point. 
Even if $\epsilon_2 \to \infty$, one would clip the weights and offer $R\epsilon_1$-DP privacy for this non-private dataset. 
In other words, this privacy comes for free!
\end{rem}

\subsection{Interpreting the Bounds on $\Le$:} 

\textbf{Only $\epsilon_2$-private dataset:} This case can be realized in different three ways: $f=0$, or $\epsilon_1=\epsilon_2$, or $\epsilon_1 = 0$\footnote{For $\epsilon_1 \to 0$, the saturation condition should be interpreted as $\epsilon_2 \leq \epsilon_1 + 8/nf\epsilon_1$}.
The first case implies $\frac{\epsilon_2}{\epsilon_1} \leq R \to \infty$ so \eqref{eq:T1A} gives an error of order $O(\frac{1}{n} + \frac{1}{(n\epsilon_2)^2})$, the minimax bound known for a homogeneous $\epsilon_2$-DP mean estimation with $n$ datapoints. 
For the second case, the same result applies as $1 = \frac{\epsilon_2}{\epsilon_1} < R$. 
In the third case, $\epsilon_2 \leq \epsilon_1 R$ so the error is of order $O(\frac{1}{n(1-f)} + \frac{1}{(n(1-f)\epsilon_2)^2})$ - again matching the minimax bound for homogeneous $\epsilon_2$-DP mean estimation with $n(1-f)$ datapoints.
There are $n(1-f)$ datapoints since the algorithm needs to be independent of the $\epsilon_1$-private data for $\epsilon_1 = 0$.

\textbf{An $\epsilon_1$-private dataset and a public dataset:} Letting $\epsilon_2 \to \infty$ implies a completely public dataset. 
Keeping $\epsilon_1$ fixed, 
since  $\frac{\epsilon_2}{\epsilon_1} \geq R$, 
\eqref{eq:T1B} yields an error of $\frac{nf\epsilon_1^2 + 8}{4n[nf\epsilon_1^2+8(1-f)]}$. 
At first glance, it behaves roughly like $\frac{1}{4n}$, corresponding to having $n$ public samples.
However, this is misleading since we care about the sharp dependence on $f,\epsilon_1$ and the factor of  $\frac{nf\epsilon_1^2 + 8}{nf\epsilon_1^2+8(1-f)}$ accounts for this, as we demonstrate next.

\textit{Tight in $f$:} consider $\epsilon_1 \to 0$, and  we get an error bound of $\frac{1}{4n(1-f)}$, which  corresponds to the known bound for having $n(1-f)$ public samples (and thus, the bound is sharp in $f$).
As a side note, depending on how we take the limits for $(\epsilon_1,\epsilon_2) \to (0,\infty)$, we may end up in case (A) in Theorem~\ref{thm:UB1} as well but the upper bound is identical for both cases for this limit.

\textit{Tight in $\epsilon_1$:} Taking $f = 1$ yields an error of $O(\frac{1}{n} + \frac{1}{(n\epsilon_1)^2})$, the minimax rate for $n$ users under homogeneous $\epsilon_1$-DP.

%% file: proofs.tex
\section{Proof Skteches} \label{sec:skt}

Detailed proofs can be found in the Appendix.

\subsection{Upper Bound}  \label{sec:skt-ub}

The randomized affine estimator
$\inprod{\vecx}{\vecw} + L(\eta)$,
where $L(\eta)$ 
is zero-mean Laplace noise with parameter $\eta$, can be shown to be $(\vecw/\eta)$-DP (see Lemma~\ref{lem:eps-af} in Appendix~\ref{A:a}).
Choosing $w_i = \epsilon_i/\Lnorm{\bm{\epsilon}}{1}$ and $\eta = 1/\Lnorm{\bm{\epsilon}}{1}$ is a possible way to satisfy the constraints - 
this suboptimal estimator proportionally weighs the datapoints based on (lack of) privacy requirement. 
If we have one datapoint with huge $\epsilon_2$ and all the other $(n-1)$ datapoints have small $\epsilon_1$, then proportional weighting will essentially use only one sample to estimate the mean and this leads to higher variance. 
Instead, we find the optimal weights $\vecw$ by optimizing the worst-case MSE.

\subsection{Lower Bound}  \label{sec:skt-lb}

We use Le Cam's method specialized to differential privacy for proving the lower bound, based on ideas from \cite{Duchi13,Duchi14,Duchi16}.
Our method is similar to \cite{Asu22} but it is more robust since it can handle arbitrarily large $\epsilon$ values, as is required for the case when we have a public dataset. Intuitively, DP restricts the variation in output probability with varying inputs which helps  bound the total-variation norm term in Le Cam's method.

%% file: experiments.tex
\begin{figure*}
    \centering
    \includegraphics[scale=0.75]{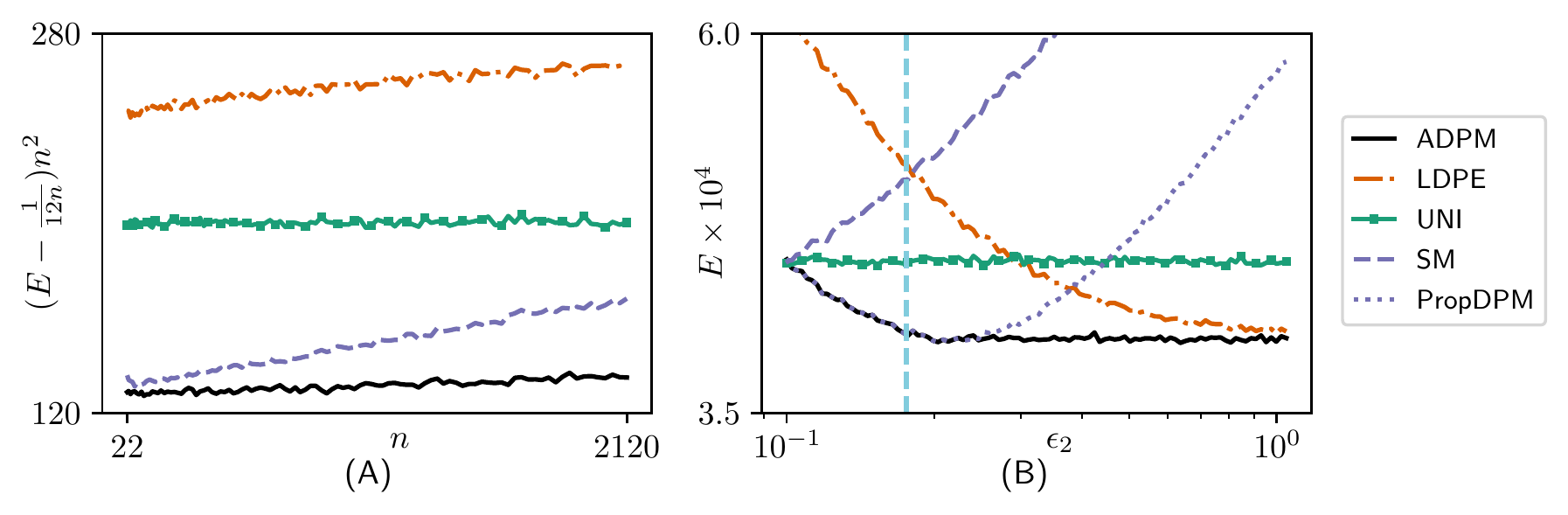}
    \vspace{-0.2in}
    \caption{
    We compare ADPM (our method) to other baseline methods in the above two graphs. FME is not plotted since its performance is an order worse than others in the above graphs; a comparison against FME is presented in Appendix~\ref{Apx:D}.  
    Subfigure (A)  plots $(E(M)-\frac{1}{12n})n^2$ vs $n$ for each algorithm, keeping $\epsilon_1 = 0.1,\ \epsilon_2 = 0.15,\ f=0.5$.  
    Subfigure (B)  plots $E(M) \times 10^4$ vs $\epsilon_2$ while keeping $\epsilon_1 = 0.1,\ f=0.7,\ n=10^3$. The vertical dashed line marks the saturation point of ADPM at $\epsilon_2 = R\epsilon_1$.
    PropDPM shows the degradation in performance of the proportional weighting scheme for larger $\epsilon_2$. 
    }
    \label{fig:compare}
    \vspace{-13pt}
\end{figure*}

\section{Experiments} \label{sec:Exp}

\subsection{Baseline Schemes}

We consider some baseline techniques for comparison and comment on why they are not optimal in HDP. 

\textbf{Uniformly enforce $\epsilon_1$-DP (UNI):} This approach offers $\epsilon_1$ privacy to all the datapoints and uses the minimax estimator, i.e., the sample mean added with Laplace noise, to get an error of $O(1/n + 1/(n\epsilon_1)^2)$. 
UNI can be arbitrarily worse than the ADPM (consider  a single low $\epsilon_1$ datapoint). 

\textbf{Sampling Mechanism (SM):} 
Based on the work of \cite{Jorg15}, let $t = \Lnorm{\bm\epsilon}{\infty}$, and sample $i$-th datapoint independently with probability $(e^{\epsilon_i}-1)/(e^t - 1)$. 
Apply homogeneous $t$-DP minimax estimator on the sub-sampled dataset.
\cite{Jorg15} proved this mechanism is $\bm\epsilon$-DP.
However, %
when one dataset is public, the SM algorithm disregards the $\epsilon_1$-private dataset.

\textbf{Local Differential Private Estimator (LDPE):} Consider the algorithm that combines the $\epsilon_1$-DP and $\epsilon_2$-DP mean estimates from the two datasets in a linear fashion. 
That is, it adds Laplace noise $L(\frac{1}{\epsilon_1nf})$ to the sample mean of the first dataset and independent Laplace noise $L(\frac{1}{\epsilon_2n(1-f)})$ to the sample mean of the second dataset, followed by optimal linear combinations of these two aggregates to minimize the mean squared error if the variance of the unknown distribution is known (see \cite{Ferrando21} for details).
We take the worst-case variance as a proxy in our problem setting.
When $\epsilon_1$ and $\epsilon_2$ are nearly the same, LDPE is worse since it adds more noise than necessary - this is a known shortcoming of the Local-DP model. 
ADPM scales better to the general case of $\bm{\epsilon}$-DP but LDPE is a decent baseline to compare it with.
Note that LDPE is optimal as well when $\epsilon_2 \to \infty$ (see Remark~\ref{rem:LDPE} in Appendix~\ref{Apx:D}).

\textbf{FME \cite{Asu22}:} For brevity, we direct the readers to \cite[Theorem 1]{Asu22} for details on the algorithm.
We refer to this algorithm as FME in the rest of this work.
One of the shortcomings of this method is it assumes $\Lnorm{\bm\epsilon}{\infty} \leq 1$ for its theoretical guarantees.
For our experiments, we still use this algorithm as it is stated for $\Lnorm{\bm\epsilon}{\infty} > 1$.
Even when $\Lnorm{\bm\epsilon}{\infty}  \leq  1$, FME may assign much smaller weights than what ADPM does to the less private dataset (Example~\ref{ex:asu} in Appendix~\ref{Apx:D}).
This might be one of the reasons why this method does not show much improvement when $\epsilon_2$ is increased (Figure~\ref{fig:JointApd} in Appendix~\ref{Apx:D}).

\textbf{Proportional DP (PropDPM):} We refer to the affine estimator with weights proportional to the $\bm\epsilon$ vector 
and appropriate Laplace noise 
as PropDPM, the shortcomings of this estimator is described in Section~\ref{sec:skt-ub}.

\noindent \textit{B. Empirical Results}

We compare  ADPM (our method) to PropDPM, LDPE, FME, SM, and UNI.
To prevent cluttering the graphs, we do not perform experiments for the stretching mechanism proposed by \cite{Alaggan17}. 
We can construct a case to demonstrate its sub-optimality since it is a biased estimator (see Example~\ref{ex:alg} in Appendix~\ref{Apx:D}).
Experiments with FME are deferred to Figure~\ref{fig:JointApd} in  Appendix~\ref{Apx:D} since it is an order of magnitude worse than other algorithms.
All simulations are averaged over 200K runs of the algorithms.

In Figure~\ref{fig:compare}(A), we plot $(\text{MSE}-\frac{1}{12n})n^2$ vs $n$ 
keeping $\epsilon_1,\epsilon_2,f$ constant for ADPM, LDPE, SM, and UNI.
We plot $(\text{MSE}-\frac{1}{12n})n^2$ to show the second-order behavior of the considered algorithms.
The simulations are run with the true underlying distribution being the uniform distribution on $\cX$.

When $n$ is small, $R$ is large, and weights proportional to $\bm\epsilon$ are optimal. 
SM algorithm does something similar so it is close to ADPM in performance.
However, since it sub-samples the data, its MSE decays slightly slower in the first order and we can see this by the upward trend in the graph.
The fact that LDPE algorithm performs worse than ADPM or UNI is not surprising since for the case considered, $\epsilon_1$ and $\epsilon_2$ are quite close so the additional noise it adds contributes to its sub-optimality.

Another insightful experiment, presented in Figure~\ref{fig:compare}(B), is to vary $\epsilon_2$ while keeping other parameters fixed and comparing the MSE.
The true underlying distribution for this experiment is the Bernoulli distribution on $\{-0.5,0.5\}$.
The curve for ADPM reinforces Theorem~\ref{thm:UB1} as there is no improvement in the MSE upon increasing $\epsilon_2$ above  $R\epsilon_1$. 
Further, PropDPM performs worse after this critical point as we expected from the discussion in Section~\ref{sec:skt-ub}.
As $\epsilon_2 \to \infty$, LDPE would achieve the optimal error 
but it does not appear to have any saturation phenomenon. 
This can be attributed to the suboptimal way of adding noise inherent to Local-DP.

 Now consider the HDP setting in its full generality of arbitrary $\bm\epsilon$.
The minimization in ADPM can be solved efficiently by modern solvers.
We consider two cases for $\bm\epsilon$ of size $10^3$ - high variance and low variance in $\bm\epsilon$.
The low variance case was obtained by uniformly sampling $\log \bm\epsilon$ in $[-3,-2]$.
Independently, the high variance case corresponds to sampling  $\log \bm\epsilon$ in $[-4,2]$. 
Keeping the sampled $\bm\epsilon$ fixed, the average of the squared errors was taken over 20K simulations under Beta$(2,3)$ distribution on $\cX$. 
The results are presented in Table~\ref{tab:het-dp}.
Unsurprisingly, UNI, PropDPM and ADPM enjoy similar  performance in the low variance regime, while diverging in the higher variance regime.

\begin{table}
\caption{Comparison of MSE for high and low variance in $\bm\epsilon$.}
\label{tab:het-dp}
\centering
\begin{tabular}{ccc}
\toprule
\textbf{Method} &
\begin{tabular}[c]{@{}c@{}} \boldmath$\log$ \textbf{MSE} \\ High $\mathsf{Var}(\bm\epsilon)$\end{tabular} &
\begin{tabular}[c]{@{}c@{}} \boldmath$\log$ \textbf{MSE} \\ Low $\mathsf{Var}(\bm\epsilon)$\end{tabular} \\
  \midrule
ADPM & -\textbf{9}.\textbf{3} & -\textbf{8}.\textbf{1} \\
PropDPM & -9.0 & -8.1 \\
LDPE & -7.2  & -1.3\\
SM & -6.5 & -7.9\\
FME & -6.2 & -6.2 \\
UNI & -5.1 & -7.1 \\
\bottomrule
\end{tabular}
\vspace{-15pt}
\end{table}

%% file: conclusion.tex
\vspace{-3pt}

\section{Conclusion} \label{sec:Con}

We study the problem of mean estimation of bounded random variables under Heterogeneous Differential Privacy and propose the ADPM algorithm.
Under HDP, when there are two groups of users with distinct privacy levels, we prove the minimax optimality of the algorithm.
Experimentally our algorithm outperforms other methods even in the general HDP setting with many distinct privacy levels.

A line of future work that we are currently working on is to prove the optimality of our algorithm in the general setting of arbitrary $\bm{\epsilon}$ vector.
The problem of mean estimation is also interesting in the unbounded setting under suitable assumptions such as sub-Gaussianity. %
Extending HDP for the multivariate case is another exciting avenue to consider.

%% file: ack.tex
\section*{Acknowledgements}

We thank Justin Singh Kang and Yigit Efe Erginbas for valuable discussions.  This work was supported in part by NSF CCF-1750430 and CCF-2007669.

%% file: AppA.tex
\section{Useful Results}\label{A:a}
Lemma~\ref{lem:eps-af} is included for completeness (see \cite{Dwork14Alg,Asu22}).
\begin{lemma}[Laplace Mechanism] \label{lem:eps-af}
The affine estimator $M_{\vecw}(\vecx) = \inprod{\vecx}{\vecw} + L(\eta)$ is $(\vecw/\eta)$-DP when $\cX = [-0.5,0.5]$.
\end{lemma}
\begin{proof}
We verify this by comparing the density of output of the algorithm on neighboring datasets. We also drop the subscript for the algorithm $M_{\vecw}$.

\begin{align}
    \frac{p(M(\vecx) = s)}{p(M(\vecx'_i) = s)} &=  \frac{\exp\{-|\inprod{\vecx}{\vecw}-s|/\eta\}}{\exp\{-|\inprod{\vecx_i'}{\vecw} - s|/\eta\}}, \\
    &\leq \exp\{|\inprod{\vecx_i'}{\vecw}-\inprod{\vecx}{\vecw}|/\eta\}, \\
    &\leq \exp\{w_i/\eta\} \label{eq:Lap},
\end{align}
where \eqref{eq:Lap} follows from the fact that $\inprod{\vecx_i'}{\vecw}$ and $\inprod{\vecx}{\vecw}$ can differ by at most $w_i$ since $w_i$ is the coefficient of the $i$-th element and $\vecx_i,\vecx_i' \in [-0.5,0.5]$.
\end{proof}

\begin{lemma} \label{lem:DP-imply}
By the definition of $\bm\epsilon$-DP in \eqref{eq:DP-def}, it follows that for all measurable sets $S \subseteq \cY$,
\begin{align} \label{eq:DP-corr}
    e^{-\epsilon_i} \pr{M(\vecx_{i}') & \in S}
     \leq \pr{M(\vecx) \in S} \\
     &\leq \begin{cases}
    e^{\epsilon_i} \pr{M(\vecx_{i}') \in S}\\
    1-e^{-\epsilon_i} + e^{-\epsilon_i}\pr{M(\vecx_{i}') \in S} \end{cases}
\end{align}
Further, it follows that
\begin{equation}
   | \pr{M(\vecx) \in S} - \pr{M(\vecx_{i}') \in S} | \leq 1-e^{-\epsilon_i} \label{eq:Delta-Bound} 
\end{equation}
\begin{proof}
Note that the DP definition also implies $e^{-\epsilon_i} \pr{M(\vecx_{i}') \in S}
     \leq \pr{M(\vecx) \in S}$.
By applying the definition with $S^C$ and combining the conditions, one obtains
\[
 \begin{rcases}
    e^{-\epsilon_i} \pr{M(\vecx_{i}') \in S}\\
    1-e^{\epsilon_i} +e^{\epsilon_i}\pr{M(\vecx_{i}') \in S}
    \end{rcases}
     \leq \pr{M(\vecx) \in S}  \]
and 
\[ \pr{M(\vecx) \in S} 
\leq \begin{cases}
    e^{\epsilon_i} \pr{M(\vecx_{i}') \in S}\\
    1-e^{-\epsilon_i} + e^{-\epsilon_i}\pr{M(\vecx_{i}') \in S}
\end{cases}\]
The condition $1-e^{\epsilon_i} +e^{\epsilon_i}\pr{M(\vecx_{i}') \in S} \leq \pr{M(\vecx) \in S}$ can be removed since $e^{-\epsilon_i} \lambda \geq 1-e^{\epsilon_i} +e^{\epsilon_i}\lambda$ for all $0 \leq \lambda \leq 1$ and non-negative $\epsilon_i$. 
The Lemma gives a much stronger bound than the straightforward DP definition when $\epsilon_i$ is large. Using \eqref{eq:DP-corr}, one can obtain \eqref{eq:Delta-Bound}.
\end{proof}
\end{lemma}

%% file: AppB.tex
\section{Upper Bound Proof}
\begin{figure*}[htbp]
  \centering
  \includegraphics[width=\textwidth]{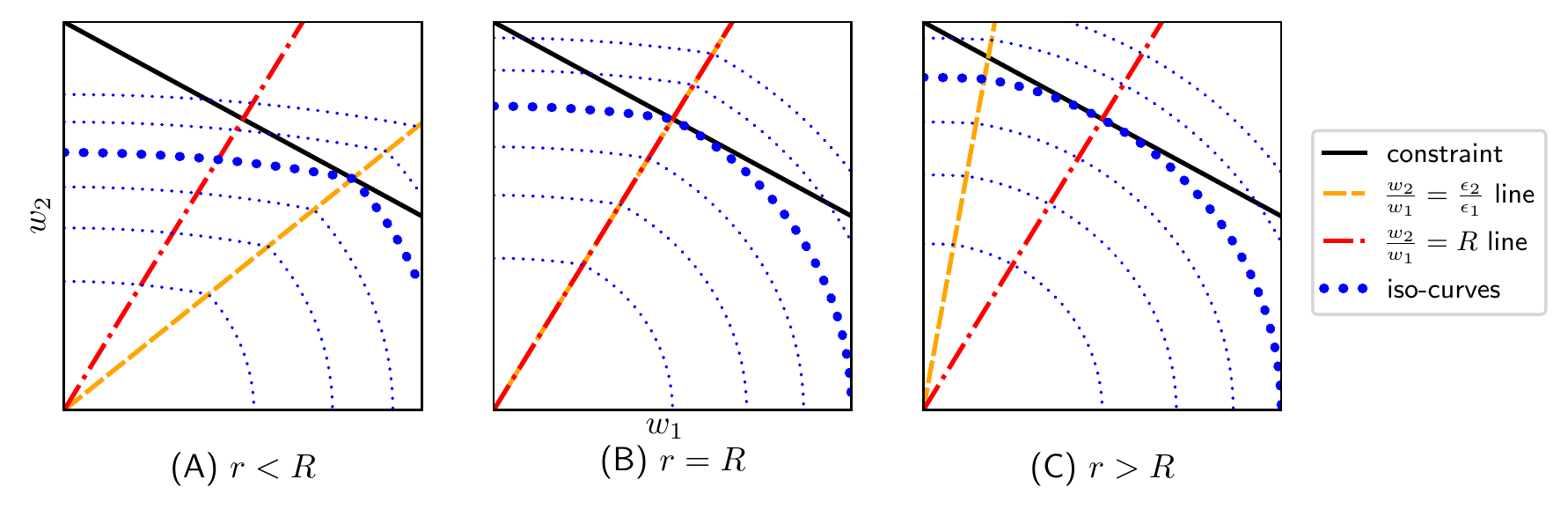}
  \caption{
    The constraint \eqref{eq:cons} is plotted in solid line while $w_2 = \epsilon_2 w_1/\epsilon_1$ and $w_2 = Rw_1$ are plotted in dash and dash-dot lines respectively for reference. 
    The isocurves for \eqref{eq:cvx-program} are plotted in dotted lines and isometric value increase away from the origin. 
    The optimal isocurve is plotted in thicker dotted lines. The graphs are plotted for $\epsilon_1 = 10^{-2},\ n=8 \times 10^4$ and $f=0.5$. 
    $\epsilon_2$ is set as $1.48 \times 10^{-2}$, $3 \times 10^{-2}$ and $1\times10^{-1}$ in (A), (B), and (C) respectively.}
  \label{fig:ineq-short}
\end{figure*}

Consider the affine estimator based on the Laplace mechanism \cite{Dwork06,Asu22} 
\begin{equation} \label{eq:Mw}
M_{\vecw}(\vecx) = \inprod{\vecx}{\vecw} + L(\eta),
\end{equation}
where $L(\eta)$ 
is zero-mean Laplace noise with parameter $\eta$.
The above estimator is $(\vecw/\eta)$-DP (element-wise division) by Lemma~\ref{lem:eps-af}. 
We impose unbiasedness along with the privacy constraint by ensuring $\vecw$ is in the $n$-dimensional standard simplex and 
$w_i \leq \epsilon_i \eta$.
The MSE of the estimator under distribtuion $P$ is given by $\var{P}\Lnorm{w}{2}^2 + 2\eta^2 \leq \Lnorm{w}{2}^2/4 + 2\eta^2$.
To minimize this, we set $\eta = \max_i w_i/\epsilon_i$.

Based on the case of two different $\epsilon$ that we consider, with some abuse of indexing, the problem of minimizing variance can be rewritten as,
\begin{align}
\text{min }  \Big\{ \frac{n(fw_1^2 + (1-f)w_2^2)}{4} + 2  &\max\Big\{\frac{w_1}{\epsilon_1},\frac{w_2}{\epsilon_2}\Big\}^2 \Big\} \label{eq:cvx-program}\\
\text{Subject to: }  fw_1 + (1-f)w_2 &= 1/n \ . \label{eq:cons}
\end{align}
In \eqref{eq:cvx-program}, $w_1$ represents the weights assigned to all the datapoints with $\epsilon_1$ privacy requirement and similarly $w_2$. 
To understand the solution to the convex program in \eqref{eq:cvx-program}, we consider the isocurves (contour plots) for 
\eqref{eq:cvx-program} and plot it along with the constraint \eqref{eq:cons} for three possible cases in Figure~\ref{fig:ineq-short}.

The solution to the convex program is presented in Lemma~\ref{lem:ub}
We summarize the conditions and optimal values in Table~\ref{tab:optimality}. 
It should be noted that we are given $n,f,\epsilon_1,\epsilon_2$ values beforehand so if the upper bound on error for this algorithm is larger than $\frac{1}{4}$, we can output $0$. 
In such a case, the minimax error will be $\frac{1}{4}$ due to the support of the random variables being $[-0.5,0.5]$. 
This proves Theorem~\ref{thm:UB1}.

\begin{lemma} \label{lem:ub}
The solution to
\begin{align}
\mathsf{min}  \ \frac{n(fw_1^2 + (1-f)w_2^2)}{4} + 2  &\max\left\{\frac{w_1}{\epsilon_1},\frac{w_2}{\epsilon_2}\right\}^2 \label{eq:cvx-program2}\\
\mathsf{Subject\ to \ \ }  fw_1 + (1-f)w_2 &= 1/n \label{eq:cons2} \ ,
\end{align}
is given by \\
(A) if $1 \leq r  \leq R$: 
$$ w_1 = \frac{\epsilon_1}{n\bar{\epsilon}};\ \ w_2 = \frac{\epsilon_2}{n\bar{\epsilon}} $$
(B) if $R \leq r$: 
$$ w_1 = \frac{1}{n[f+(1-f)R]}; \ \ w_2 = \frac{R}{n[f+(1-f)R]} $$
\end{lemma}
\begin{proof}
The isocurves (contour plots) for 
\eqref{eq:cvx-program2} and the constraint curve in \eqref{eq:cons2} are plotted in Figure~\ref{fig:ineq-short}. 
As we increase $r$, the ratio of epsilons, the point where the optimal solution $(w_1,w_2)$ switches from being proportional to $(\epsilon_1,\epsilon_2)$ to being clipped to a lower value, we have $w_1/\epsilon_1 = w_2/\epsilon_2$ as well as the following tangent condition
\begin{align}
\frac{dw_2}{dw_1} &= \frac{-f}{1-f}, \label{eq:der-cons}\\
\frac{n}{2}\left[ fw_1 + (1-f)w_2 \frac{dw_2}{dw_1} \right] &+ 4w_1/\epsilon_1^2 = 0. \label{eq:der-cur}
\end{align}
 
\eqref{eq:der-cons} corresponds to the slope of the constraint curve while \eqref{eq:der-cur} gives a condition on the slope of the objective function. Substitution  \eqref{eq:der-cons} in \eqref{eq:der-cur} and using $w_1/\epsilon_1 = w_2/\epsilon_2$, we get 
 \begin{equation}
    R  = r = \frac{\epsilon_2}{\epsilon_1} = 1 + \frac{8}{\epsilon_1^2 nf}
 \end{equation}

  Thus, for $r < R$, we an obtain the optimal weights by using the fact that optimal weights are in ratio of the epsilons, as shown in Figure~\ref{fig:ineq-short}. 
 The optimal weights are $w_i = \frac{\epsilon_i}{n\bar{\epsilon}}$ for $i=1,2$. 
 The minimum objective value is given as $\frac{\overline{\epsilon^2}}{4 n \bar{\epsilon}^2} + \frac{2}{(n\bar{\epsilon})^2}$, where $\bar{\epsilon} = f\epsilon_1 + (1-f)\epsilon_2$ and $\overline{\epsilon^2} = f\epsilon_1^2 + (1-f)\epsilon_2^2$.

If $r > R$,  we can obtain the weights by the tangency condition of \eqref{eq:der-cons} and \eqref{eq:der-cur} to get $w_1 = \frac{1}{n[f+(1-f)R]},\ w_2 =  \frac{R}{n[f+(1-f)R]}$. The corresponding minimum objective value is given by $\frac{R}{4n[f+(1-f)R]}$.
\end{proof}

%% file: AppC.tex
\section{Lower Bound Proof} \label{apx:LB}
We use the shorthand $\bm{L}$ to represent $\Le$ in this section.
\begin{lemma} \label{lem:lb}
We have the following lower bound on $\bm{L}$,
\begin{align}
\bm{L} &\gtrsim \frac{1}{6n} \wedge \frac{1}{4}, \\
\bm{L} &\gtrsim  \frac{1}{ \left( 4nf\epsilon_1 + 4\sqrt{ 8n(1-f) } \right)^2 } \wedge \frac{1}{4},  \\
\bm{L} &\gtrsim  \frac{f(R-1)}{n[f + (1-f)r]^2} \wedge \frac{1}{4}.
\end{align}
\end{lemma}
\begin{proof}
    Let $P_1$, $P_2$ be two distributions in $\cP$ and let $M$ be any $\bm\epsilon$-DP estimator of the mean, then denote the distribution of $M(\vecX)$ with $\vecX \sim P_i^n$ as $Q_i$ for $i=1,2$.
    In other words, $Q_i$ is a distribution over $\cY$ and $Q_i(A) = \bbP_{\vecX \sim P_i^n}\{M(\vecX) \in A\}$.

Consider the distribution $P_1$ which is $0.5$ with probability $\frac{1+\delta}{2}$ and -0.5 with probability $\frac{1-\delta}{2}$. Similarly, $P_2$ is $0.5$ with probability $\frac{1-\delta}{2}$ and -0.5 with probability $\frac{1+\delta}{2}$. \\

In this case, $\mu_{P_1} = \delta/2$ and $\mu_{P_2} = -\delta/2$.
Further, $\Lnorm{P_1 - P_2}{TV} = \delta$ and $\kl{P_1}{P_2} \leq 3 \delta^2 $ (for $\delta \in [0,0.5]$). 
Define $\gamma = \frac{1}{2}\left| \mu_{P_1} - \mu_{P_2} \right| = \delta/2$
, then, Le Cam's method specialized to differential privacy setting (see \cite{Duchi13,Duchi14,Duchi16}) yields the lower bound
\begin{equation} \label{eq:LeC}
\bm{L} \geq \frac{\gamma^2}{2}(1 - \Lnorm{Q_1 - Q_2}{TV}).
\end{equation}
Treating the privacy requirement as the vector $\bm\epsilon$, 
using Lemma~\ref{lem:TV-bound}, and $1-x\leq e^{-x}\, \forall x\geq 0$, we obtain 
\begin{equation} \label{eq:QB}
\Lnorm{Q_1 - Q_2}{TV} \leq 2\delta \sum_{i=1}^k \epsilon_i + \delta\sqrt{\frac{3(n-k)}{2}} \quad \forall k \in \{0,\ldots,n\}. 
\end{equation}
Note that \eqref{eq:QB} holds for arbitrarily large $\epsilon_i$ values and degrades gracefully as compared to the $e^{\epsilon_i} - 1$ bound obtained in \cite[Lemma 3]{Asu22}. 
We could achieve this due to the stronger bound we derive in Lemma~\ref{lem:DP-imply} and Lemma~\ref{lem:TV-bound}. In particular, this allows us to deal with the general case when one of the datasets is public.

In \eqref{eq:QB}, arrange the $\epsilon_i$ in ascending order so that the first $nf$ values are $\epsilon_1$ followed by $\epsilon_2$. Using $k=0,nf,n$ in \eqref{eq:QB}, get
\begin{align}
\Lnorm{Q_1 - Q_2}{TV} &\leq \delta\sqrt{\frac{3n}{2}} ,\label{eq:QB1} \\
\Lnorm{Q_1 - Q_2}{TV} &\leq 2\delta nf\epsilon_1 + \delta\sqrt{\frac{3n(1-f)}{2}} \\
&\leq 2\delta nf\epsilon_1 + 2\delta\sqrt{8n(1-f)}, \label{eq:QB2} \\
\Lnorm{Q_1 - Q_2}{TV} &\leq 2\delta [nf\epsilon_1 + n(1-f)\epsilon_2 ],\label{eq:QB3}
\end{align} 
respectively. Using \eqref{eq:QB1}, \eqref{eq:QB2}, and \eqref{eq:QB3} in \eqref{eq:LeC}, we obtain 
\begin{align}
\bm{L} &\geq \frac{\delta^2}{8}\left(1 - \delta\sqrt{\frac{3n}{2}}\right), \label{eq:L-1}\\
\bm{L} &\geq \frac{\delta^2}{8}\left(1 - \delta \left(2nf\epsilon_1 + 2\sqrt{8n(1-f)}\right)\right), \label{eq:L-2}\\
\bm{L} &\geq \frac{\delta^2}{8}\left(1 - 2 \delta n[f\epsilon_1 + (1-f)\epsilon_2] \right) , \label{eq:L-3}
\end{align}
respectively. Setting $\delta = \frac{1}{\sqrt{6n}} \wedge 0.5$, $\delta = \frac{1}{4nf\epsilon_1 + 4\sqrt{8n(1-f)}} \wedge 0.5$, and $\delta = \frac{1}{4n[f\epsilon_1 + (1-f)\epsilon_2]} \wedge 0.5$, in \eqref{eq:L-1}, \eqref{eq:L-2}, and \eqref{eq:L-3}, respectively, to get the claimed lower bounds on $\Le$.
\end{proof}

For clarity, we work with the universal constants in Lemma~\ref{lem:lb}.

\begin{align}
\bm{L} &\geq \frac{1}{16} \left\{ \frac{1}{6n} \wedge \frac{1}{4} \right\} ,  \label{eq:A}\\
\bm{L} &\geq  \frac{1}{16} \left\{ \frac{1}{ \left( 4nf\epsilon_1 + 4\sqrt{ 8n(1-f) } \right)^2 } \wedge \frac{1}{4} \right\},  \label{eq:tB} \\
\bm{L} &\geq  \frac{1}{16} \left\{ \frac{1}{16n^2[f\epsilon_1 + (1-f)\epsilon_2]^2} \wedge \frac{1}{4} \right\}  \\
&= \frac{1}{16} \left\{ \frac{f(R-1)}{128n[f + (1-f)r]^2} \wedge \frac{1}{4}  \right\} \label{eq:c-p1}\\
&\geq \frac{1}{512} \left\{ \frac{f(R-1)}{4n[f + (1-f)r]^2} \wedge \frac{1}{4}  \right\}\label{eq:C},
\end{align}

where we used $n\epsilon_1^2 = 8/f(R-1)$ in \eqref{eq:c-p1}. Using $(a+b)^2 \leq 2(a^2 + b^2)$ in \eqref{eq:tB}
and $f\leq1$,
\begin{align} \label{eq:B}
\bm{L} &\geq \frac{1}{16} \left\{  \frac{1}{32\left[ n^2 f \epsilon_1^2 + 8n(1-f) \right]} \wedge \frac{1}{4} \right\}
\end{align}

Take the convex combination of \eqref{eq:A}$\times \frac{6}{6+256f}$ and \eqref{eq:B}$\times \frac{256f}{6+256f}$ and use $6+256f \leq 262$ to get
\begin{align} 
\bm{L} &\geq  \frac{1}{16} \left\{ \frac{nf\epsilon_1^2 + 8}{262n[ n f \epsilon_1^2 + 8(1-f)] } \wedge \frac{1}{4} \right\} \\
&=  \frac{1}{16} \left\{ \frac{R}{262n[ f + R(1-f)] } \wedge \frac{1}{4} \right\} \\
&\geq  \frac{1}{1048} \left\{ \frac{R}{4n[ f + R(1-f)] } \wedge \frac{1}{4} \right\}
\label{eq:lb-sat}
\end{align}

\eqref{eq:lb-sat} proves Theorem~\ref{thm:LB1}(B). Since this lower bound is independent of $r$, it also holds in the regime $r \leq R$ but we show a stronger lower bound in this regime.

\subsection{\boldmath$1 \leq r\leq R$ Regime}

Let $L_1 = \frac{1}{6n}$, $L_2 = \frac{R}{4n[f+(1-f)R]}$, and $L_3(r) = \frac{f(R-1)}{4n[f+r(1-f)]^2}$. 
$L_1$, $L_2$, and $L_3(r)$ are implicit functions of $n,f,\epsilon_1$ while $L_3(r)$ also depends on $r$. We restrict our attention to the case when $r \in [1,R]$. 
Recall that $R = 1 + \frac{8}{nf\epsilon_1^2}$ so the range of $r$ under consideration  is also an implicit function of $n,f,\epsilon_1$.
By \eqref{eq:A}, \eqref{eq:lb-sat}, and \eqref{eq:C}  we have, $\bm{L} \gtrsim L_1 \wedge \frac{1}{4}, L_2 \wedge \frac{1}{4}, L_3(r) \wedge \frac{1}{4}$ respectively. 

By taking convex combination,
\begin{equation}
    \bm{L} \geq \frac{1}{1560} \left\{  (L_2 + L_3(r)) \wedge \frac{1}{4}  \right\}. \label{eq:L23-bound}
\end{equation}
Let $U(r) = \frac{fR+(1-f)r^2}{4n[f+(1-f)r]^2}$. $U(r)$ is the upper bound on error we obtained in the regime $1 \leq r \leq R$. 
Then, $U(r) - L_3(r) = \frac{f+(1-f)r^2}{4n[f+(1-f)r]^2}$.
Oberve the derivative with respect to $r$
\begin{equation}
    U'(r) - L_3'(r) = \frac{(1-f)f(r-1)}{2n[f+r(1-f)]^3} \geq 0 \ \ \forall r \geq 1.
\end{equation}
Thus, $U(r) - L_3(r)$ is the largest at $r=R$ and is given by 
\begin{equation}
    U(R) - L_3(R) = \frac{f + R^2(1-f)}{4n[f+R(1-f)]^2}.
\end{equation}
Further,
\begin{equation}
    U(R) - L_3(R) - L_2= \frac{f(1-R)}{4n[f+R(1-f)]^2} \leq 0.
\end{equation}
Thus, $U(r) \leq L_3(r) + L_2$ $\forall r \in [1,R]$. Using this in \eqref{eq:L23-bound}, we obtain,
\begin{equation}
    \bm{L} \geq \frac{1}{1560} \left\{ U(r) \wedge \frac{1}{4} \right\} \ \ \forall \ r\in [1,R],
\end{equation}
which proves Theorem~\ref{thm:LB1}(A).

\begin{lemma} \label{lem:TV-bound}
For any $k \in \{0,1,\ldots,n\}$,
\begin{align}
    \Lnorm{Q_1 - Q_2}{TV} \leq 2\Lnorm{P_1 & - P_2}{TV} \sum_{i=1}^k(1-e^{-\epsilon_i}) \\ 
    &+ \sqrt{\frac{n-k}{2}\kl{P_1}{P_2}}.
\end{align}
 
\begin{proof}
We use a method similar to \cite{Asu22} to prove this but obtain stronger results due to Lemma~\ref{lem:DP-imply}. Let $\tilde Q$ be the distribution of the output of $\bm\epsilon$-DP estimator $M(\cdot)$ when the input is dataset $\vecX$ draw from the product distribution $P_1^kP_2^{n-k}$. 
By triangle inequality,
\begin{equation}
   \Lnorm{Q_1 - Q_2}{TV} \leq \Lnorm{Q_1 - \tilde Q}{TV} + \Lnorm{\tilde Q - Q_2}{TV}. 
\end{equation}

By Data Processing Inequality and Pinsker's inequality, 
\begin{equation}
   \Lnorm{Q_1 - \tilde Q}{TV} \leq  \Lnorm{P_1^n - P_1^kP_2^{n-k}}{TV} \leq \sqrt{\frac{n-k}{2}\kl{P_1}{P_2}}. 
\end{equation}

\begin{figure*}[htbp]
    \centering
    \includegraphics{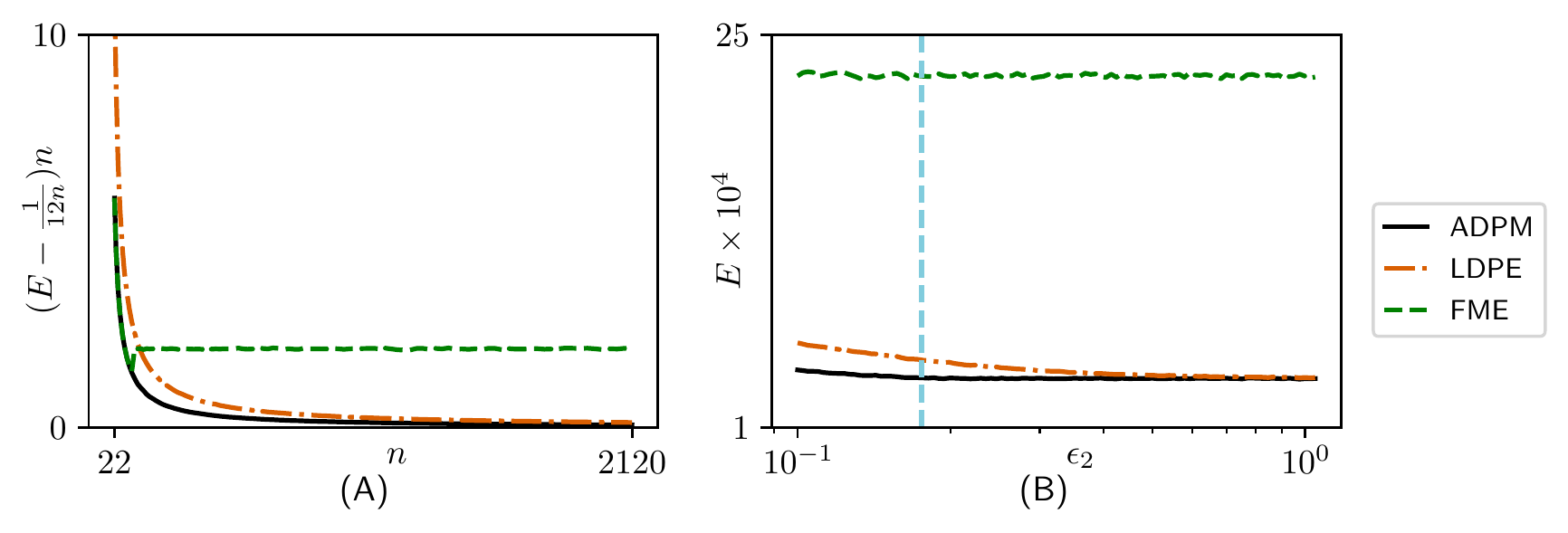}
    \caption{We compare FME with other algorithms in terms of $( E(M)-1/12n)n$ vs. $n$ and $E(M) \times 10^4$ vs  $\epsilon_2$. In (B), the vertical dashed line denotes the value of $\epsilon_2 = R\epsilon_1$.}
    \label{fig:JointApd}
\end{figure*}

\begin{figure}
    \centering
    \includegraphics[width=0.45\textwidth]{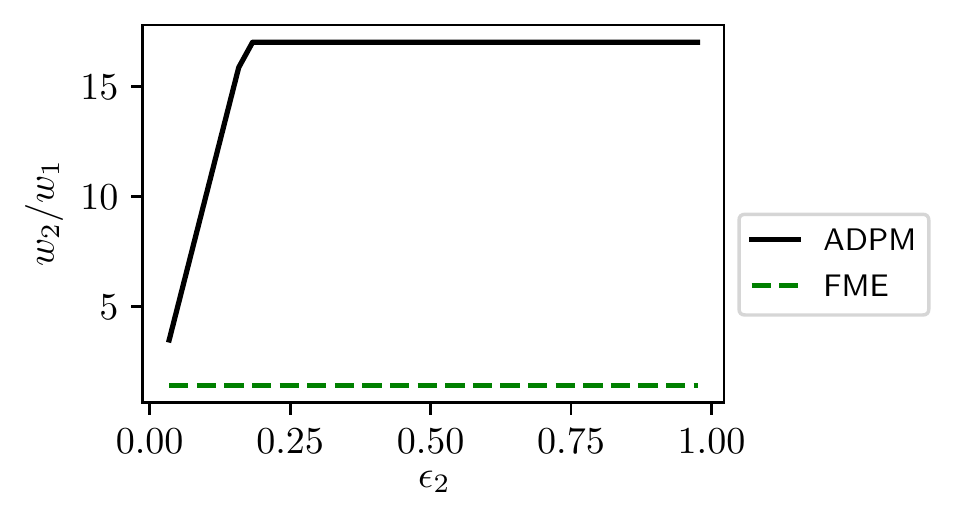}
    \caption{The ratio of weights $w_2/w_1$ is plotted as $\epsilon_2$ is increased for ADPM and FME.}
    \label{fig:FME-SAT}
\end{figure}

Next, consider the term $\Lnorm{\tilde Q - Q_2}{TV}$,
\begin{align}
    |\tilde Q(A) &- Q_2(A)| =  | \bbE_{\vecX \sim P_1^kP_2^{n-k}}\pr{M(\vecX) \in A}  \\
    &\quad \quad \quad \quad \quad - \bbE_{\vecX \sim P_2^{n}}\pr{M(\vecX) \in A} | \\
    &= \Big| \sum_{i=1}^k \big( \bbE_{\vecX \sim P_1^iP_2^{n-i}}\pr{M(\vecX) \in A} \\ &\quad \quad - \bbE_{\vecX \sim P_1^{i-1}P_2^{n-i+1}}\pr{M(\vecX) \in A} \big) \Big| \\
    &\leq \sum_{i=1}^k \big|\bbE_{\vecX \sim P_1^iP_2^{n-i}}\pr{M(\vecX) \in A} \\ 
    &\quad \quad \quad - \bbE_{\vecX \sim P_1^{i-1}P_2^{n-i+1}}\pr{M(\vecX) \in A} \big| .
\end{align}
Let $\vecX_i'$ denote another dataset which differs from $\vecX$ at only the $i$-th index and this index can have any arbitrary value independent of $\vecX$. Then, note that 
\begin{align}
\bbE_{\vecX \sim P_1^iP_2^{n-i}}&\pr{M(\vecX_i') \in A} \\
&- \bbE_{\vecX \sim P_1^{i-1}P_2^{n-i+1}}\pr{M(\vecX_i') \in A} = 0 \ \ \ a.s.
\end{align}
Thus,
\begin{align}
    &|\tilde Q(A) - Q_2(A)| \leq  \\
    &\sum_{i=1}^k \Big|\bbE_{\vecX \sim P_1^iP_2^{n-i}}\big[\pr{M(\vecX) \in A}-\pr{M(\vecX_i') \in A}\big]\\
    &-\bbE_{\vecX \sim P_1^{i-1}P_2^{n-i+1}}[\pr{M(\vecX) \in A}-\pr{M(\vecX_i') \in A}]  \Big|. \label{eq:ExpDiff}
\end{align}
Let $\vecX_{-i}$ denote the random vector $\vecX$ except at the $i$-th position, i.e., $\vecX_{-i} \sim P_1^{i-1}P_2^{n-i}$ means that elements of $\vecX$ from position $1$ to position $i-1$ are I.I.D. $P_1$ while those in positions $i+1$ to $n$ are I.I.D. $P_2$.

Therefore, we get,
\begin{align}
    &|\tilde Q(A) - Q_2(A)| \leq \\
    &\sum_{i=1}^k \Bigg|\bbE_{\vecX_{-i} \sim P_1^{i-1}P_2^{n-i}}\Big[\bbE_{X_i \sim P_1}[\pr{M(\vecX) \in A} \\
    &\hspace{130pt}  -\pr{M(\vecX_i') \in A}]  \\
     &\quad\quad -  \bbE_{X_i \sim P_2}[\pr{M(\vecX) \in A}-\pr{M(\vecX_i') \in A}]\Big]  \Bigg| \\ 
    &\leq  \sum_{i=1}^k \bbE_{\vecX_{-i} \sim P_1^{i-1}P_2^{n-i}}\Big| \bbE_{X_i \sim P_1}[\pr{M(\vecX) \in A}\\ 
    &\hspace{130pt} -\pr{M(\vecX_i') \in A}]  \\
     &\quad \quad - \bbE_{X_i \sim P_2}[\pr{M(\vecX) \in A}-\pr{M(\vecX_i') \in A}]\Big| \\
    &\leq  \sum_{i=1}^k \bbE_{\vecX_{-i} \sim P_1^{i-1}P_2^{n-i}}[ 2(1-e^{-\epsilon_i})\Lnorm{P_1 - P_2}{TV}]  \label{eq:Jump} \\
    &= 2\Lnorm{P_1 - P_2}{TV}\sum_{i=1}^k(1-e^{-\epsilon_i}).
\end{align}

In \eqref{eq:Jump}, we used Lemma~\ref{lem:DP-imply} and the fact that for a bounded function $|f(x)| \leq C$, we have $\left|E_{X\sim P_1}[f(X)] - E_{X\sim P_2}[f(X)] \right| \leq 2\Lnorm{P_1 - P_2}{TV}C$.

\end{proof}
\end{lemma}

%% file: AppD.tex
\section{Additional Material} \label{Apx:D}

Figure~\ref{fig:JointApd}(A) plots $( E(M)-1/12n)n$ vs. $n$ for $\epsilon_1 = 0.1$, $\epsilon_2 = 0.15$, and $f=0.5$.
Note that in Figure~\ref{fig:compare}(A), we had plotted $( E(M)-1/12n)n^2$ instead but we plot $(E(M)-1/12n)n$ since FME degrades poorly with $n$.
Figure~\ref{fig:JointApd}(B) plots $E(M) \times 10^4$ vs $\epsilon_2$ while keeping $\epsilon_1 = 0.1,\ f=0.7,\ n=10^3$.
In both figures, we can see that ADPM performs better than FME and LDPE, and FME performs an order of magnitude worse than the other methods.

An example is presented in Example~\ref{ex:asu} to demonstrate that the FME method can fail to properly identify the saturation phenomenon we see in ADPM algorithm.

\begin{ex}[Saturation in FME] \label{ex:asu}
    Consider $\epsilon_1 = 0.01$, $n=10^4$, $f=0.5$.
    We vary $\epsilon_2$ and plot the ratio of $w_2/w_1$ obtained by FME and ADPM in Figure~\ref{fig:FME-SAT}.
    As Figure~\ref{fig:FME-SAT} shows, at times, FME may not be able to properly saturate the weights as $\epsilon_2$ is changed and maintains the ratio $w_2/w_1 = \sqrt{2}$.
\end{ex}

\begin{ex}[Method of \cite{Alaggan17}] \label{ex:alg}
   The shrinkage matrix-based mechanism proposed in \cite{Alaggan17} considers $\epsilon_{max} = \max \bm\epsilon$ and then scales $i$-th datapoint $x_i$ with privacy requirement $\epsilon_i$ to $\epsilon_i x_i/\epsilon_{max}$. After such scaling, one can apply any homogeneous $\epsilon_{max}$-DP algorithm on the dataset.
   This method is not suitable for mean estimation since it adds a bias. Consider the distribution which has a probability mass of $1$ at $0.5$. Setting $\epsilon_1 = 0.01$, $\epsilon_2 = 0.1$, $f=0.5$,  the MSE is given by $0.05 + 200/n^2$, which can not be made arbitrarily small for large $n$.
\end{ex}

\begin{rem}[Optimality of LDPE with public dataset] \label{rem:LDPE}
For the case of two datasets, when one dataset is public, LDPE can achieve the same MSE as ADPM, i.e., in Figure~\ref{fig:compare}(B), LDPE asymptotically achieves the same error as ADPM. 
To see this, consider the worst-case MSE of the two datasets separately using Local-DP: the $\epsilon_1$-private dataset has an error of $E_1 \leq 1/4nf + 2/(nf\epsilon_1)^2$ while the public dataset has an error of $E_2 \leq 1/4n(1-f)$. 
The optimal linear combination of these two Local-DP estimates weighs the estimate of the $\epsilon_1$-private dataset by $E_2/(E_1 + E_2)$ and the public dataset by $E_1/(E_1 + E_2)$.
The MSE of the linear combination is upper bounded by $E_1E_2/(E_1 + E_2)$, which on simplifying gives $\frac{R}{4n[f+(1-f)R]}$.
\end{rem}

%% file: main.bbl
\begin{thebibliography}{10}
\providecommand{\url}[1]{#1}
\csname url@samestyle\endcsname
\providecommand{\newblock}{\relax}
\providecommand{\bibinfo}[2]{#2}
\providecommand{\BIBentrySTDinterwordspacing}{\spaceskip=0pt\relax}
\providecommand{\BIBentryALTinterwordstretchfactor}{4}
\providecommand{\BIBentryALTinterwordspacing}{\spaceskip=\fontdimen2\font plus
\BIBentryALTinterwordstretchfactor\fontdimen3\font minus
  \fontdimen4\font\relax}
\providecommand{\BIBforeignlanguage}[2]{{%
\expandafter\ifx\csname l@#1\endcsname\relax
\typeout{** WARNING: IEEEtran.bst: No hyphenation pattern has been}%
\typeout{** loaded for the language `#1'. Using the pattern for}%
\typeout{** the default language instead.}%
\else
\language=\csname l@#1\endcsname
\fi
#2}}
\providecommand{\BIBdecl}{\relax}
\BIBdecl

\bibitem{Hoffman69}
L.~J. Hoffman, ``Computers and privacy: A survey,'' \emph{ACM Computing
  Surveys}, vol.~1, no.~2, p. 85–103, June 1969.

\bibitem{Survey00}
R.~Agrawal and R.~Srikant, ``Privacy-preserving data mining,'' in
  \emph{Proceedings of the 2000 ACM SIGMOD international conference on
  Management of data}, 2000, pp. 439--450.

\bibitem{Rao18}
P.~Ram Mohan~Rao, S.~Murali~Krishna, and A.~Siva~Kumar, ``Privacy preservation
  techniques in big data analytics: a survey,'' \emph{Journal of Big Data},
  vol.~5, pp. 1--12, 2018.

\bibitem{GDPR}
EU, ``Regulation ({EU}) 2016/679 of the {E}uropean {P}arliament and of the
  {C}ouncil of 27 april 2016 on the protection of natural persons with regard
  to the processing of personal data and on the free movement of such data, and
  repealing directive 95/46/ec ({General Data Protection Regulation}),'' pp.
  1--88, May 2016.

\bibitem{CCPA}
CA, ``California {C}onsumer {P}rivacy {A}ct ({CCPA}),'' \emph{Office of the
  Attorney General, California Department of Justice}, 2018.

\bibitem{DW06}
C.~Dwork, K.~Kenthapadi, F.~McSherry, I.~Mironov, and M.~Naor, ``Our data,
  ourselves: Privacy via distributed noise generation,'' in \emph{Annual
  international conference on the theory and applications of cryptographic
  techniques}.\hskip 1em plus 0.5em minus 0.4em\relax Springer, 2006.

\bibitem{Dwork06}
C.~Dwork, F.~McSherry, K.~Nissim, and A.~Smith, ``Calibrating noise to
  sensitivity in private data analysis,'' in \emph{Theory of cryptography
  conference}.\hskip 1em plus 0.5em minus 0.4em\relax Springer, 2006.

\bibitem{Mironov17}
I.~Mironov, ``R{\'e}nyi differential privacy,'' in \emph{2017 IEEE 30th
  computer security foundations symposium (CSF)}.\hskip 1em plus 0.5em minus
  0.4em\relax IEEE, 2017.

\bibitem{Dwork16}
C.~Dwork and G.~N. Rothblum, ``Concentrated differential privacy,'' \emph{arXiv
  preprint arXiv:1603.01887}, 2016.

\bibitem{Bun16}
M.~Bun and T.~Steinke, ``Concentrated differential privacy: Simplifications,
  extensions, and lower bounds,'' in \emph{Theory of Cryptography
  Conference}.\hskip 1em plus 0.5em minus 0.4em\relax Springer, 2016.

\bibitem{Wang20}
T.~Wang, X.~Zhang, J.~Feng, and X.~Yang, ``A comprehensive survey on local
  differential privacy toward data statistics and analysis,'' \emph{Sensors},
  2020.

\bibitem{Cai19}
T.~T. Cai, Y.~Wang, and L.~Zhang, ``{The cost of privacy: Optimal rates of
  convergence for parameter estimation with differential privacy},'' \emph{The
  Annals of Statistics}, 2021.

\bibitem{Kotsogiannis20}
I.~Kotsogiannis, S.~Doudalis, S.~Haney, A.~Machanavajjhala, and S.~Mehrotra,
  ``One-sided differential privacy,'' in \emph{2020 IEEE 36th International
  Conference on Data Engineering (ICDE)}.\hskip 1em plus 0.5em minus
  0.4em\relax IEEE, 2020.

\bibitem{Kamath19}
G.~Kamath, J.~Li, V.~Singhal, and J.~Ullman, ``Privately learning
  high-dimensional distributions,'' in \emph{Conference on Learning
  Theory}.\hskip 1em plus 0.5em minus 0.4em\relax PMLR, 2019.

\bibitem{Duchi14}
R.~F. Barber and J.~C. Duchi, ``Privacy and statistical risk: Formalisms and
  minimax bounds,'' 2014.

\bibitem{Hopkins22}
S.~B. Hopkins, G.~Kamath, and M.~Majid, ``Efficient mean estimation with pure
  differential privacy via a sum-of-squares exponential mechanism,'' in
  \emph{Proceedings of the 54th Annual ACM SIGACT Symposium on Theory of
  Computing}, 2022.

\bibitem{Kamath20P}
G.~Kamath and J.~Ullman, ``A primer on private statistics,'' \emph{arXiv
  preprint arXiv:2005.00010}, 2020.

\bibitem{Kasiviswanathan11}
S.~P. Kasiviswanathan, H.~K. Lee, K.~Nissim, S.~Raskhodnikova, and A.~Smith,
  ``What can we learn privately?'' \emph{SIAM Journal on Computing}, 2011.

\bibitem{Duchi16}
J.~C. Duchi, M.~J. Wainwright, and M.~I. Jordan, ``Minimax optimal procedures
  for locally private estimation,'' \emph{Journal of the American Statistical
  Association}, 2016.

\bibitem{Bassily20}
R.~Bassily, A.~Cheu, S.~Moran, A.~Nikolov, J.~Ullman, and S.~Wu, ``Private
  query release assisted by public data,'' in \emph{International Conference on
  Machine Learning}.\hskip 1em plus 0.5em minus 0.4em\relax PMLR, 2020.

\bibitem{Bassily20-learn}
R.~Bassily, S.~Moran, and A.~Nandi, ``Learning from mixtures of private and
  public populations,'' \emph{Advances in Neural Information Processing
  Systems}, 2020.

\bibitem{Liu21}
T.~Liu, G.~Vietri, T.~Steinke, J.~Ullman, and S.~Wu, ``Leveraging public data
  for practical private query release,'' in \emph{International Conference on
  Machine Learning}.\hskip 1em plus 0.5em minus 0.4em\relax PMLR, 2021.

\bibitem{Alon19}
N.~Alon, R.~Bassily, and S.~Moran, ``Limits of private learning with access to
  public data,'' \emph{Advances in neural information processing systems},
  2019.

\bibitem{Nandi20}
A.~Nandi and R.~Bassily, ``Privately answering classification queries in the
  agnostic pac model,'' in \emph{Algorithmic Learning Theory}.\hskip 1em plus
  0.5em minus 0.4em\relax PMLR, 2020.

\bibitem{Kairouz21}
P.~Kairouz, M.~R. Diaz, K.~Rush, and A.~Thakurta, ``({N}early) dimension
  independent private erm with adagrad rates via publicly estimated
  subspaces,'' in \emph{Conference on Learning Theory}.\hskip 1em plus 0.5em
  minus 0.4em\relax PMLR, 2021.

\bibitem{Amid22}
E.~Amid, A.~Ganesh, R.~Mathews, S.~Ramaswamy, S.~Song, T.~Steinke, V.~M.
  Suriyakumar, O.~Thakkar, and A.~Thakurta, ``Public data-assisted mirror
  descent for private model training,'' in \emph{International Conference on
  Machine Learning}.\hskip 1em plus 0.5em minus 0.4em\relax PMLR, 2022.

\bibitem{Wang19}
D.~Wang, H.~Zhang, M.~Gaboardi, and J.~Xu, ``Estimating smooth glm in
  non-interactive local differential privacy model with public unlabeled
  data,'' in \emph{Algorithmic Learning Theory}.\hskip 1em plus 0.5em minus
  0.4em\relax PMLR, 2021.

\bibitem{Kamath22}
A.~Bie, G.~Kamath, and V.~Singhal, ``Private estimation with public data,'' in
  \emph{Advances in Neural Information Processing Systems}, 2022.

\bibitem{Liu21Pj}
J.~Liu, J.~Lou, L.~Xiong, J.~Liu, and X.~Meng, ``Projected federated averaging
  with heterogeneous differential privacy,'' \emph{Proceedings of the VLDB
  Endowment}, 2021.

\bibitem{Alaggan17}
M.~Alaggan, S.~Gambs, and A.-M. Kermarrec, ``Heterogeneous differential
  privacy,'' \emph{Journal of Privacy and Confidentiality}, 2017.

\bibitem{Li17Par}
H.~Li, L.~Xiong, Z.~Ji, and X.~Jiang, ``Partitioning-based mechanisms under
  personalized differential privacy,'' in \emph{Pacific-asia conference on
  knowledge discovery and data mining}.\hskip 1em plus 0.5em minus 0.4em\relax
  Springer, 2017.

\bibitem{Jorg15}
Z.~Jorgensen, T.~Yu, and G.~Cormode, ``Conservative or liberal? personalized
  differential privacy,'' in \emph{2015 IEEE 31St international conference on
  data engineering}, 2015.

\bibitem{Ferrando21}
C.~Ferrando, J.~Gillenwater, and A.~Kulesza, ``Combining public and private
  data,'' in \emph{NeurIPS 2021 Workshop Privacy in Machine Learning}, 2021.

\bibitem{Niu20}
B.~Niu, Y.~Chen, B.~Wang, J.~Cao, and F.~Li, ``Utility-aware exponential
  mechanism for personalized differential privacy,'' in \emph{2020 IEEE
  Wireless Communications and Networking Conference (WCNC)}, 2020.

\bibitem{McSherry07}
F.~McSherry and K.~Talwar, ``Mechanism design via differential privacy,'' in
  \emph{48th Annual IEEE Symposium on Foundations of Computer Science
  (FOCS'07)}.\hskip 1em plus 0.5em minus 0.4em\relax IEEE, 2007.

\bibitem{Li17CF}
Y.~Li, S.~Liu, J.~Wang, and M.~Liu, ``A local-clustering-based personalized
  differential privacy framework for user-based collaborative filtering,'' in
  \emph{International Conference on Database Systems for Advanced
  Applications}.\hskip 1em plus 0.5em minus 0.4em\relax Springer, 2017.

\bibitem{Zhang19}
S.~Zhang, L.~Liu, Z.~Chen, and H.~Zhong, ``Probabilistic matrix factorization
  with personalized differential privacy,'' \emph{Knowledge-Based Systems},
  2019.

\bibitem{Chen16}
R.~Chen, H.~Li, A.~K. Qin, S.~P. Kasiviswanathan, and H.~Jin, ``Private spatial
  data aggregation in the local setting,'' in \emph{2016 IEEE 32nd
  International Conference on Data Engineering (ICDE)}, 2016.

\bibitem{Cum22Mean}
R.~Cummings, V.~Feldman, A.~McMillan, and K.~Talwar, ``Mean estimation with
  user-level privacy under data heterogeneity,'' \emph{Advances in Neural
  Information Processing Systems}, vol.~35, pp. 29\,139--29\,151, 2022.

\bibitem{Torkamani22}
S.~Torkamani, J.~B. Ebrahimi, P.~Sadeghi, R.~G. D’Oliveira, and
  M.~M{\'e}dard, ``Heterogeneous differential privacy via graphs,'' in
  \emph{2022 IEEE International Symposium on Information Theory (ISIT)}, 2022.

\bibitem{Chatzikokolakis13}
K.~Chatzikokolakis, M.~E. Andr{\'e}s, N.~E. Bordenabe, and C.~Palamidessi,
  ``Broadening the scope of differential privacy using metrics,'' in
  \emph{International Symposium on Privacy Enhancing Technologies
  Symposium}.\hskip 1em plus 0.5em minus 0.4em\relax Springer, 2013.

\bibitem{Avent17}
B.~Avent, A.~Korolova, D.~Zeber, T.~Hovden, and B.~Livshits, ``{BLENDER}:
  Enabling local search with a hybrid differential privacy model,'' in
  \emph{26th USENIX Security Symposium}, 2017.

\bibitem{Ave20}
B.~Avent, Y.~Dubey, and A.~Korolova, ``The power of the hybrid model for mean
  estimation,'' \emph{Proceedings on Privacy Enhancing Technologies}, vol.
  2020, pp. 48--68, 10 2020.

\bibitem{Beimel19}
A.~Beimel, A.~Korolova, K.~Nissim, O.~Sheffet, and U.~Stemmer, ``{The Power of
  Synergy in Differential Privacy: Combining a Small Curator with Local
  Randomizers},'' in \emph{1st Conference on Information-Theoretic Cryptography
  (ITC 2020)}, 2020.

\bibitem{Asu22}
A.~Fallah, A.~Makhdoumi, A.~Malekian, and A.~Ozdaglar, ``Optimal and
  differentially private data acquisition: Central and local mechanisms,'' in
  \emph{Proceedings of the 23rd ACM Conference on Economics and
  Computation}.\hskip 1em plus 0.5em minus 0.4em\relax Association for
  Computing Machinery, 2022.

\bibitem{Duchi13}
J.~Duchi, M.~Jordan, and M.~Wainwright, ``Local privacy and minimax bounds:
  Sharp rates for probability estimation,'' \emph{Adv. Neural Inform. Process.
  Syst}, 2013.

\bibitem{Dwork14Alg}
C.~Dwork, A.~Roth \emph{et~al.}, ``The algorithmic foundations of differential
  privacy,'' \emph{Foundations and Trends in Theoretical Computer Science},
  vol.~9, no. 3--4, pp. 211--407, 2014.

\end{thebibliography}
